\documentclass[aps,twocolumn,pra,superscriptaddress,showpacs,tightenlines]{revtex4-1}
\usepackage{amssymb}
\usepackage{amsmath}
\usepackage{graphicx}
\usepackage{epsfig}
\usepackage{subfigure}
\usepackage{amsfonts}

\begin{document}

\title{Zeno-anti-Zeno crossover via external fields in a one-dimensional coupled-cavity waveguide}

\author{Lan Zhou}
\affiliation{Key Laboratory of Low-Dimensional Quantum Structures
and Quantum Control of Ministry of Education, and Department of
Physics, Hunan Normal University, Changsha 410081, China}
\author{Le-Man Kuang}
\affiliation{Key Laboratory of Low-Dimensional Quantum Structures
and Quantum Control of Ministry of Education, and Department of
Physics, Hunan Normal University, Changsha 410081, China}

\begin{abstract}
We have studied a hybrid system of a one-dimensional coupled-cavity
waveguide with a two-level system inside, which subject to a
external periodical field. Using the extended Hilbert space
formalism, the time-dependent Hamiltonian is reduced into an
equivalent time-independent one. Via computing the Floquet-Green's
function, the Zeno-anti-Zeno crossover is controlled by the driven
intensity and frequency, and the detuning between the cavity and the
two-level system.
\end{abstract}
\maketitle \narrowtext 
\narrowtext

\section{\label{Sec:1}Introduction}

Inspired by modern microfabrication technology in photonic crystals\cite%
{JAP75(94)4753,PRB54(96)007837,OL24(99)711,APL84(04)161,Ozbay2000}, optical
microcavities\cite{Vahala2003}, superconducting devices \cite%
{Makhlin2001,YouNori2005,Clarke2008}, and the realization of the quantum
regime in the interaction of atomic-like structures and quantized
electromagnetic modes in those system\cite%
{Armani03,n431(04)162,n431(04)159,Hennessy07,Trupke07}, considerable
attention has been attracted to a family of models for coupled arrays of
atom-cavity systems, since they offer a fascinating combination of condensed
matter physics and quantum optics. Typically, these models are formed by an
array of cavities with each cavity containing one or more atoms\cite%
{PlenioRV08}, where photons hop between the cavities. Recently, due to the
potential use for building a quantum switch for routing single-photons in
quantum network, systems with one or two atoms inside the array of cavities
have been extensively studied to revealed the intriguing features of photon
transport in low dimensional environments\cite%
{PRL101(08)100501,PRA78(08)053806,PRA78(08)063827,PRA80(09)062109,PRA79(09)063847}%
. It is found that the switch is formed by the interference between the
spontaneous emission from atoms and the propagating modes in the
one-dimensional (1D) continuum. Therefore, the search for a controllable
switch is to finding a way to change the spontaneous emission of atoms.

The spontaneous emission results from the inevitable interaction of the
atomic system with external influences. Decoherence of a quantum system,
with a variety of couplings to a reservoir, has been investigated
extensively in theory\cite{ZurekPT91,OSAQO,GardinerQN}. Basically, there are
three ways to suppress or modify the rate of quantum transitions in a
system. One is to engineer the state of the reservoir, as well as the form
of the system-reservoir coupling\cite{ERN403,Raimond01,Dorner02}. Obviously,
the coupled-cavity waveguide (CRW) meets this condition for its advantage of
addressability of individual sites, extremely high controllability, and the
great degree of flexibility in their geometric design. Two is to involve the
quantum inference between multiple transition pathways of internal states,
for example, the electromagnetically induced transparency technique\cite%
{HarrisPT,YLEIT69,ZLEIT76}. The third way is to applies a succession of
short and strong pulses, or measurement to the quantum system\cite%
{BBcontrol,pulsePRL06,pulsePRL07,pulsePRL08,mesureN00,ZLPRA06}.

Atoms have been particularly interested for acting as a quantum node in the
extended communication networks and scalable computational devices,
specially artificial atoms. Atoms in a time-varying field has been
investigated long time before. However, the multiphoton resonance and
quantum interference have been experimentally demonstrated in a strongly
driven artificial atom\cite{Oliver05} until recently, which is important to
superconducting approach of quantum computation, for instance, decreasing
the time required for each gate operation\cite%
{Vion02,Urbina02,MooijS03,Tsai06}.

Floquet theory is developed by Floquet centuries ago, it is a theory about
the solutions of linear differential equations with periodic coefficients%
\cite{Floquet}. Later, it is applied to the two-level system (TLS) with the
time-dependent problem, which is discussed by Autler and Townes\cite{Autler}%
. And then such kind of the periodic time-dependent problem is reformulated
as an equivalent time-independent infinite-dimensional Floquet matrix\cite%
{Shirley}, which is done by introducing the composite Hilbert space of
square integrable and time-periodic wave functions. Now Floquet formalism is
used as a theoretical tool to investigate time dependent phenomena\cite%
{SIchupr}. In this paper, we study a two-level system (e.g. a flux qubit)
interacting with a cavity which together with other cavities constructs a
one-dimensional (1D) coupled-cavity waveguide (CRW). The CRW is modeled as a
linear chain of sites with the nearest-neighbor interaction. Obviously, the
dynamic of the TLS is irreversible, i.e. once the TLS is initially in its
excited state, it never returns to its initial state spontaneously.
Therefore, the CRW with a TLS inside is a typical system with a discrete
state coupled to continua of states, which means the TLS is subject to
decay. In order to control the decay rate of the TLS, an external periodical
field is applied to the TLS through diagonal coupling. It is well known that
periodic coherent pulses can either inhibit or accelerate the decay into its
reservoir, however, such modification is based on off-diagonal
time-dependent couplings between the TLS and the external field. In a atomic
experiment, the off-diagonal couplings is caused by a transverse field
perpendicular to the polarizing field, which is used to polarize the atomic
spins. Then the spins of atoms and photon are aligned. However, diagonal
couplings between the driving field and the TLS means that the driving field
is parallel to the polarizing field\cite{Wilson}. Here, the spontaneous
emission of a TLS is studied, the Zeno-anti-Zeno crossover is investigated
via Floquet formulation.

The paper is organized as follows. To set our consideration, in Sec.~\ref%
{Sec:2}, we state the Hamiltonian for 1D CRW with a TLS inside one of the
cavity. in Sec.~\ref{Sec:3}, we first give a brief outline of the Floquet
theorem, then present the Floquet representation of the system we consider.
In Sec.~\ref{Sec:4}, we calculate the amplitude for the TLS in its excited
state in the one quantum subspace, and the quantum Zeno and anti-Zeno
crossover is investigated. Conclusions are summarized at the end of the
paper.

\section{\label{Sec:2} Description of the model.}

Consider a TLS subject to an harmonically transverse driving with frequency $%
\nu $ and intensity $A$. The lower and upper eigenstates of the two-level
system are described by notation $|e\rangle $ and $\,|g\rangle $
respectively, which is separated by energy $\Omega $ without external
fields. Such kind of TLSs can been experimentally realized using
superconducting circuits\cite{Oliver05,Wilson}. The Hamiltonian describing
the TLS is%
\begin{equation}
H_{A}=\left( \Omega +A\cos \nu t\right) \sigma _{z}/2,  \label{1zaz-1}
\end{equation}%
where $\sigma _{z}$ describes the atomic inversion. $\Omega +A\cos \nu t$ is
the energy splitting. The TLS interacts with a quantized electromagnetic
field of a 1D waveguide, which is constructed by coupling of cavities in an
array. Due to the overlap of the spatial profile of the cavity modes, photon
hops between neighbouring cavity.
\begin{figure}[tbp]
\includegraphics[bb=83 446 514 667, width=8.5 cm,clip]{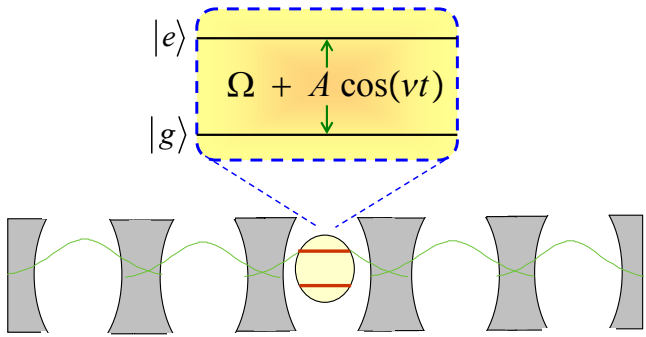}
\caption{(Color online) Schematic illustration of the model, where a two
level system is inside a 1D coupled cavity waveguide. The two level system
is driven by external forces that are periodic in time with period $\protect%
\nu $.}
\label{fig2:1}
\end{figure}
Introducing the creation and annihilation operators of the cavity modes, $%
a_{j}^{\dag }$ and $a_{j}$, the CRW is modeled as a linear chain of sites
with the nearest-neighbor interaction. The Hamiltonian of the 1D CRW can be
written as%
\begin{equation}
H_{W}=\sum_{j}\omega _{c}a_{j}^{\dag }a_{j}-\xi \sum_{j}\left( a_{j+1}^{\dag
}a_{j}+h.c.\right) ,  \label{1zaz-2}
\end{equation}%
where $\xi $ is the hopping energy between adjacent cavities and $\omega
_{c} $ is the eigenfrequency of each cavity. The TLS is inside one of the
cavity in the 1D CRW, which is labeled as the zeroth cavity. In the rotating
wave approximation (RWA), the interaction between the TLS and the zeroth
cavity is described by a Jaynes-Cummings Hamiltonian%
\begin{equation}
H_{I}=g\left( \sigma _{-}a_{0}^{\dag }+h.c.\right) .  \label{1zaz-3}
\end{equation}%
The operators $\sigma _{+}$ and $\sigma _{-}$ are the usual raising and
lowering operators of the TLS. The Hamiltonian governing the system is
\begin{equation}
H=H_{A}+H_{W}+H_{I}.  \label{1zaz-4}
\end{equation}%
By employing the Fourier transformation%
\begin{equation}
a_{j}=\frac{1}{\sqrt{N}}\sum_{k}e^{ikj}a_{k},  \label{1zaz-5}
\end{equation}%
the Hamiltonian $H$ in the Hilbert space of configuration is now given by in
the momentum space%
\begin{eqnarray}
H &=&\sum_{k}\varepsilon _{k}a_{k}^{\dag }a_{k}+\frac{1}{2}\left[ \Omega
+A\cos \left( \nu t\right) \right] \sigma _{z}  \label{1zaz-6} \\
&&+\frac{g}{\sqrt{N}}\sum_{k}\left( a_{k}^{\dag }\sigma _{-}+h.c.\right) ,
\notag
\end{eqnarray}%
where the dispersion relation%
\begin{equation}
\varepsilon _{k}=\omega _{c}-2\xi \cos k  \label{1zaz-7}
\end{equation}%
describes an energy band of width $4\xi $ (the lattice constant is assumed
to be unity). The second-quantization operators $a_{k}/a_{k}^{\dag }$
annihilate/create one photon in the kth mode of the 1D CRW. The periodical
boundary condition is used to obtain the Hamiltonian $H$ in Eq.(\ref{1zaz-6}%
).

It can be found that the number of quanta, which is defined by operator $%
\mathcal{N}=\sum_{k}a_{k}^{\dag }a_{k}+\sigma _{z}$, is conserved in this
system, i.e. $\mathcal{N}$ commutes with Hamiltonian $H$. Therefore, if we
have one quantum in the initial state, the state vector evolves restrictedly
in the one-quantum space. We introduce the states $\left\vert \bar{k}%
\right\rangle =a_{k}^{\dag }\left\vert 0g\right\rangle $ which describes
that there are one excitation in the $k$th mode of the CRW while the TLS
stay in its ground state, and $\left\vert \bar{e}\right\rangle =\left\vert
0e\right\rangle $ which denotes that the TLS has been flipped to its excited
state while the CRW is in the vacuum state. The orthonormal basis set $%
\left\{ \left\vert \bar{k}\right\rangle ,\left\vert \bar{e}\right\rangle
\right\} $ spans the one-quantum space. Therefore Hamiltonian $H$ is
rewritted as%
\begin{eqnarray}
H &=&\frac{1}{2}\left[ \Omega +A\cos \left( \nu t\right) \right] \left(
\left\vert \bar{e}\right\rangle \left\langle \bar{e}\right\vert
-\sum_{k}\left\vert \bar{k}\right\rangle \left\langle \bar{k}\right\vert
\right)  \label{1zaz-8} \\
&&+\sum_{k}\varepsilon _{k}\left\vert \bar{k}\right\rangle \left\langle \bar{%
k}\right\vert +\frac{g}{\sqrt{N}}\sum_{\bar{k}}\left( \left\vert \bar{k}%
\right\rangle \left\langle \bar{e}\right\vert +h.c.\right)  \notag
\end{eqnarray}%
in the one-quantum subspace. Hamiltonian $H$ in Eq.(\ref{1zaz-8}) describes
a single discrete state is coupled to a continuum when the periodical-driven
field is absent, which means the discrete state is subject to decay.
Consequently, the excited state of the TLS is an unstable state.

\section{\label{Sec:3}Floquet formulation for driven atom inside a 1D
waveguide.}

Since the external driving field is strictly periodic in time, Hamiltonian
in Eq.(\ref{1zaz-8}) is a periodic function in time, i.e. $H\left( t\right)
=H\left( t+T\right) $ with $T=2\pi /\nu $ being the period. To study our
problem, it is necessary to consider solutions of Schr\"{o}dinger equation
with a time periodic Hamiltonian. In this section, the theoretical tools
that we used is the Floquet representation of quantum-mechanical systems. A
brief outline of this method has been given in appendix.

We now employ the Floquet-state nomenclature\cite{SIchupr} $\left\vert
\alpha n\right\rangle =\left\vert \alpha \right\rangle \otimes \left\vert
n\right\rangle $, where $n$ is the Fourier index runing from $-\infty $ to $%
\infty $, $\alpha =\bar{e},\bar{k}$ is the system index. We first deal with
Hamiltonian $H_{A}$ in Eq.(\ref{1zaz-1}). In the Floquet space, the time
parameter is another degree of freedom of the system. Hence, the periodical
function $\cos \left( \nu t\right) $ is an operator with the expression%
\begin{equation}
\cos \left( \nu t\right) =\frac{1}{2}\sum_{n}\left( \left\vert
n+1\right\rangle \left\langle n\right\vert +h.c.\right) \text{.}
\label{rep-10}
\end{equation}%
In the one quantum subspace, Hamiltonian $H_{A}-i\partial _{t}$ can be
separated into two segments $H_{e}+H_{g}$ with
\begin{subequations}
\label{rep-11}
\begin{eqnarray}
H_{e} &=&\sum_{n}\left( \frac{\Omega }{2}+n\nu \right) \left\vert \bar{e}%
n\right\rangle \left\langle \bar{e}n\right\vert \\
&&+\sum_{n}\frac{A}{4}\left( \left\vert \bar{e}n+1\right\rangle \left\langle
n\right\vert +h.c.\right)  \notag \\
H_{g} &=&\sum_{kn}\left( -\frac{\Omega }{2}+n\nu \right) \left\vert \bar{k}%
n\right\rangle \left\langle \bar{k}n\right\vert \\
&&-\sum_{kn}\frac{A}{4}\left( \left\vert \bar{k}n+1\right\rangle
\left\langle \bar{k}n\right\vert +h.c.\right)  \notag
\end{eqnarray}%
It can be diagonalized by the following transform
\end{subequations}
\begin{subequations}
\label{rep-12}
\begin{align}
\left\vert \bar{e}n\right\rangle & =\sum_{m}J_{n-m}\left( -\chi /2\right)
\left\vert \bar{e}\phi _{m}\right\rangle \\
\left\vert \bar{k}n\right\rangle & =\sum_{m}J_{n-m}\left( \chi /2\right)
\left\vert \bar{k}\phi _{m}\right\rangle
\end{align}%
where $J_{n}\left( x\right) $ is the Bessel function of the first kind and $%
\chi =A/\nu $. In terms of states in Eq.(\ref{rep-12}) we reduce the
solution of a periodic time-dependent Hamiltonian $H$ to the problem of
diagonalizing the time-independent Floquet Hamiltonian $H_{F}=H_{0}+H_{1}$
with
\end{subequations}
\begin{subequations}
\label{rep-13}
\begin{eqnarray}
H_{0} &=&\sum_{m}E_{m}^{\bar{e}}\left\vert \bar{e}\phi _{m}\right\rangle
\left\langle \bar{e}\phi _{m}\right\vert +\sum_{km}E_{m}^{\bar{k}}\left\vert
\bar{k}\phi _{m}\right\rangle \left\langle \bar{k}\phi _{m}\right\vert \\
H_{1} &=&\sum_{kmm^{\prime }}\frac{gJ_{m-m^{\prime }}\left( \chi \right) }{%
\sqrt{N}}\left( \left\vert \bar{k}\phi _{m}\right\rangle \left\langle \bar{e}%
\phi _{m^{\prime }}\right\vert +h.c.\right) .
\end{eqnarray}%
Here, $E_{m}^{\bar{e}}$ ($E_{m}^{\bar{k}}$) is the eigenvalue of $H_{e}$($%
H_{g}$) with
\end{subequations}
\begin{subequations}
\label{rep-14}
\begin{eqnarray}
E_{m}^{\bar{e}} &=&\Omega /2+m\nu \text{,} \\
E_{m}^{\bar{k}} &=&\varepsilon _{k}-\Omega /2+m\nu \text{.}
\end{eqnarray}%
The eigenvectors of Hamiltonian $H_{e}$ and $H_{g}$ are coupled via the
nonzero coupling strength $g$. Figure~\ref{fig3:1} shows the energy diagram
of Hamiltonian $H_{e}$ and $H_{g}$ in Eq.(\ref{rep-11}) when $2\xi <\nu $.
\begin{figure}[tbp]
\includegraphics[bb=106 410 460 683, width=8.5 cm,clip]{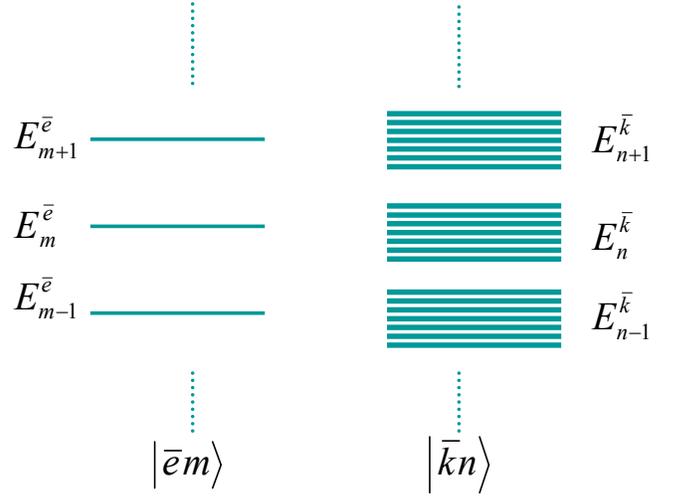}
\caption{(Color online) Schematic Quasienergy diagram of the Hamiltonian $%
H_{e}$ (a) and $H_{g}$ (b) in Eq.(\protect\ref{rep-11}) under the condition $%
2\protect\xi <\protect\nu $.}
\end{figure}
When the eigenvalues $E_{m}^{\bar{e}}$ and $E_{n}^{\bar{k}}$ are close to
each other, i.e. $E_{m}^{\bar{e}}\approx E_{n}^{\bar{k}}$, Floquet
Hamiltonian $H_{F}$ is reduced to the following form
\end{subequations}
\begin{equation}
H_{R}=H_{R0}+H_{R1}  \label{rep-15}
\end{equation}%
where
\begin{subequations}
\label{rep-16}
\begin{eqnarray}
H_{R0} &=&E_{m}^{\bar{e}}\left\vert \bar{e}\phi _{m}\right\rangle
\left\langle \bar{e}\phi _{m}\right\vert +\sum_{k}E_{n}^{\bar{k}}\left\vert
\bar{k}\phi _{n}\right\rangle \left\langle \bar{k}\phi _{n}\right\vert \\
H_{R1} &=&\sum_{k}\frac{gJ_{n-m}\left( \chi \right) }{\sqrt{N}}\left(
\left\vert \bar{k}\phi _{n}\right\rangle \left\langle \bar{e}\phi
_{m}\right\vert +h.c.\right)
\end{eqnarray}%
This approximation seems equivalent to the rotating wave approximation (RWA)
which is traditionally used in the field of atomic field to neglect the
counter-rotating term of the harmonic driving, however this approximation is
different from the RWA, which is valid only for amplitudes of the driving
field small compared to the energy difference between the atomic states and
breaks down in the strong field.

The above discussion shows that if the TLS is initially in its excited
state, it will emit a photon as a result of the interaction with the
radiation field of the CRW, photons will gain or loss energy quantum $\hbar
\omega $ due to the periodical modulation. Hence the state of the emitted
photon is characterized by the set of energy $E_{q}=\Omega -q\hbar \omega $
with $q=n-m$. Once $E_{q}$ equals to the energy $\varepsilon _{k}$ of the
CRW, photons go to the CRW. However, when ratio of the intensity to the
frequency of the modulation is equal to the roots of the Bessel function $%
J_{q}\left( \chi \right) $, this process is prevented due to the decoupling
of the TLS and CRW as one can see in Eq.(\ref{rep-16}b). When the modulation
is absent, the index $q$ vanishes (i.e. $q=0$). The behavior of the emitted
photon is determined by whether $\Omega $ is equal to $\varepsilon _{k}$ or
not. Here, we only give a intuitively discussion, more details will present
in the next section.

\section{Zeno-anti-Zeno crossover}

Time evolution of standard quantum theory assumes two principles: the
continuous unitary evolution without measurement, and the projective
measurement. Quantum Zeno effect (QZE) or anti-Zeno effect (AZE) is a
phenomenon related to projective measurements, which says that repeated
observations prolong or shorten of a lifetime of an unstable state. However,
slowdown or speedup of the decay of an unstable state is also possible
without measurements or observations\cite{Peres,wineland}.

We now investigate the lifetime of the periodically driven TLS in
interaction with the CRW when the hopping energy and the driven frequency
are chosen such that $2\xi <\nu $. An excited state of the TLS in one
quantum subspace of this system will evolves into a superposition of itself
and the states in which the atom is unexcited and has released a photon into
the CRW. With the half width of the band smaller than the driven frequency,
the dynamic of the TLS is major governed by Hamiltonian in Eq.(\ref{rep-15})
with $m=0$. In terms of the Green function, the probability for finding an
initial excited TLS still in the excited state reads $P_{e}=\left\vert \oint
dEe^{-iEt}C_{e}\left( E\right) \right\vert ^{2}$. Here we have defined
\end{subequations}
\begin{equation}
C_{e}\left( E\right) =\left\langle e\phi _{0}\right\vert \left(
E-H_{R}\right) ^{-1}\left\vert e\phi _{0}\right\rangle  \label{ZE-01}
\end{equation}%
which is the Fourier-Laplace transform of the amplitude $C_{e}\left(
t\right) $ for the TLS in its excited state at arbitary time. Since the
number of the cavities in the CRW is larger, the coupling between the TLS
and the CRW is small. Therfore, Hamiltonian $H_{R0}$ can be regarded as the
unperturbed part, while Hamiltonian $H_{R1}$ is treated as a perturbation
part. Via Dyson's equation, we compute the Floquet-Green's functions up to
the second nonvanishing order in the coupling strength $g/\sqrt{N}$. Then
the amplitude reads
\begin{equation}
C_{e}\left( E\right) =\frac{1}{E-E_{0}^{\bar{e}}}\left[ 1+\frac{\tilde{g}%
_{n}\left( E\right) }{E-E_{0}^{\bar{e}}}\right]  \label{ZE-02}
\end{equation}%
where $g_{n}\left( E\right) $ is the Fourier--Laplace transform of $%
g_{n}\left( t\right) e^{-i\left( \omega -\Omega /2+n\nu \right) \tau }$
\begin{eqnarray}
\tilde{g}_{n}\left( E\right) &=&\int_{0}^{\infty }dtg_{n}\left( \tau \right)
e^{-i\left( \omega -\Omega /2+n\nu \right) \tau }e^{iEt-\eta t}
\label{ZE-03} \\
&=&\sum_{k}\frac{g^{2}}{N}\frac{J_{n}^{2}\left( \chi \right) }{E-E_{n}^{\bar{%
k}}}.  \notag
\end{eqnarray}%
The inverse Fourier--Laplace transform of $g\left( E\right) $ yields the
memory function
\begin{equation}
g_{n}\left( t\right) =\frac{g^{2}}{N}J_{n}^{2}\left( \chi \right)
\sum_{k}e^{it2\xi \cos k}  \label{ZE-04}
\end{equation}%
or reservoir response function\cite{mesureN00,ZLPRA06}, which depends on the
quasiexcitation in the $N$ modes of the CRW and characterizes the spectrum
of the reservior. Comparing with the memory function $\Phi \left( t\right) $
in the absence of modulation, an extra factor $J_{n}^{2}\left( \chi \right) $
has been involved. Obviously, by setting $n=0$ and $A=0$, $g_{n}\left(
t\right) $ reduces to $\Phi \left( t\right) $.

The inverse Fourier-Laplace transform of Eq.(\ref{ZE-02}) yields the time
evolution of the amplitude $C_{e}\left( t\right) $%
\begin{equation}
C_{e}=e^{iE_{0}^{\bar{e}}t}\left[ 1-t\int_{0}^{t}d\tau \left( 1-\frac{\tau }{%
t}\right) g_{n}\left( \tau \right) e^{-i\left( \Delta +n\nu \right) \tau }%
\right] ,  \label{ZE-05}
\end{equation}%
where detuning $\Delta =\omega _{c}-\Omega $. The probability for finding
the TLS in its excited state reads
\begin{equation}
P_{e}\simeq \exp \left( -Rt\right) .  \label{ZE-06}
\end{equation}%
Here the decay rate is the overlap of the modulation spectrum $f_{n}\left(
\omega \right) $ and the reservoir coupling spectrum $g_{n}\left( \omega
\right) $%
\begin{equation}
R=2\pi \int_{-\infty }^{+\infty }d\omega f_{n}\left( \omega \right)
g_{n}\left( \omega \right) ,  \label{ZE-07}
\end{equation}%
where $f_{n}\left( \omega \right) $ and $g_{n}\left( \omega \right) $ is the
Fourier transform of functions $f_{n}\left( \tau \right) $ and $g_{n}\left(
\tau \right) $. The function $f_{n}\left( \tau \right) $ is defined as
\begin{equation}
f_{n}\left( \tau \right) =\left( 1-\frac{\tau }{t}\right) e^{-i\left( \Delta
+n\nu \right) t}\Theta \left( t-\tau \right) ,  \label{ZE-08}
\end{equation}%
where $\Theta \left( x\right) $ is the Heaviside unit step function, i.e., $%
\Theta \left( x\right) =1$ for $x\geq 0$, and $\Theta \left( x\right) =0$
for $x<0$. Comparing with the form factor induced by the frequently
measurement, an extra factor $e^{-in\nu t}$ has been introduced by the
modulation. However one can find that when the modulation is absent ($n=0$),
the modulation spectrum reduces to the measurement-induced level-broadening
function in Ref.\cite{mesureN00}.

The expression of functions $f_{n}\left( \tau \right) $ in Eq.(\ref{ZE-08})
and $g_{n}\left( \tau \right) $ in Eq.(\ref{ZE-04}) allows us to calculate
the decay rate as%
\begin{equation}
R=\frac{tg^{2}}{N}J_{n}^{2}\left( \chi \right) \sum_{k}\sin c^{2}\frac{%
\left( \Delta -2\xi \cos k+n\nu \right) t}{2},  \label{ZE-09}
\end{equation}%
where $\sin c=\sin x/x$. It shows that the decay rate $R$ is determined by:
1) the parameters of the driving field, i.e. the driven intensity $A$ and
frequency $\nu $; 2) the detuning $\Delta $ between the cavity and the TLS;
3) the modulation time $t$; 4) the number $N$ of cavities in the 1D
waveguide. But only frequency $\nu $, intensity $A$, and the detuning $%
\Delta $ can be adjusted experimentally.

We first consider the long-time dynamics of the periodically driven TLS. As
time $t\rightarrow \infty $, the decay rate in Eq. (\ref{ZE-09}) is a sum of
Dirac delta functions%
\begin{equation}
R=\frac{g^{2}}{N}J_{n}^{2}\left( \chi \right) \sum_{k}\delta \left( \Delta
-2\xi \cos k+n\nu \right)  \label{ZE-10}
\end{equation}%
Eq.(\ref{ZE-10}) shows that depending on whether the matching condition
\begin{equation}
\Delta -2\xi \cos k+n\nu =0  \label{ZE-11}
\end{equation}%
is satisfied, the TLS is either (i) frozen to its initial excited state or
(ii) the TLS moves to the lower state and stays in it for ever. Case (ii)
appears at the matching condition, but case (i) occurs when the transition
energy of the TLS is out of resonance with the $n$th energy band of the CRW.
When the number of cavities in the CRW is infinity, the states in the
reservoir are continuum, one can replace the sums over $k$ by integrals.
Therefore, the properties of CRW are described by the reservoir spectral
density
\begin{eqnarray}
\rho \left( \omega \right) &=&N^{-1}\sum_{k}\delta \left( \omega +2\xi \cos
k\right)  \label{ZE-12} \\
&=&\left\{
\begin{array}{c}
0\text{ \ \ \ \ \ \ \ \ \ \ \ }2\xi <\left\vert \omega \right\vert \\
\infty \text{ \ \ \ \ \ \ \ \ \ \ \ }2\xi =\left\vert \omega \right\vert \\
\frac{2/\pi }{\sqrt{4\xi ^{2}-\omega ^{2}}}\text{ \ \ }2\xi >\left\vert
\omega \right\vert%
\end{array}%
\right. .  \notag
\end{eqnarray}
And the Fourier transformation of the reservoir response function reads
\begin{equation}
g_{n}\left( \omega \right) =g^{2}J_{n}^{2}\left( A/\nu \right) \rho \left(
\omega \right) .  \label{ZE-13}
\end{equation}%
The behavior of the TLS is determined by whether the transition energy of
the TLS is inside or outside the energy band of the CRW. When $\varepsilon
_{k=0}\leq \Omega -n\nu \leq \varepsilon _{k=\pi }$, the TLS is in its
ground state and the single quantum stays in the modes of the CRW. The decay
rate reads $R=2\pi g_{n}\left( \Delta +n\nu \right) $. It is the extension
of the golden rule rate to the case of a time-dependent coupling. When the
TLS is out of resonance with the CRW, the TLS remains in its excited state.
However, all the above matching conditions doesn't matter when the ratio of
intensity $A$ to frequency $\nu $ is chosen such that $J_{n}\left( A/\nu
\right) =0$. And at those points where the driving intensity $A$ and
frequency $\nu $ satisfy $J_{n}\left( A/\nu \right) =0$, the decay is
completely suppressed. Actually, the occurrence of the complete suppression
at $J_{n}\left( A/\nu \right) =0$ is caused by the decoupling between the
TLS and the CRW. Notice that the subscript $n$ of the Bessel function is
determined by the ratio of the energy difference between the CRW and the TLS
to the frequency $\nu $, and the zeroes of the Bessel function $J_{n}\left(
x\right) $ with different $n$ appear in different argument $x$. Therefore,
one can switch on or off the coupling between the TLS and the CRW by
adjusting the detuning $\Delta $ for a given intensity $A$ and $\nu $ which
satisfy $J_{n}\left( A/\nu \right) =0$.
\begin{figure}[tbp]
\includegraphics[width=8 cm]{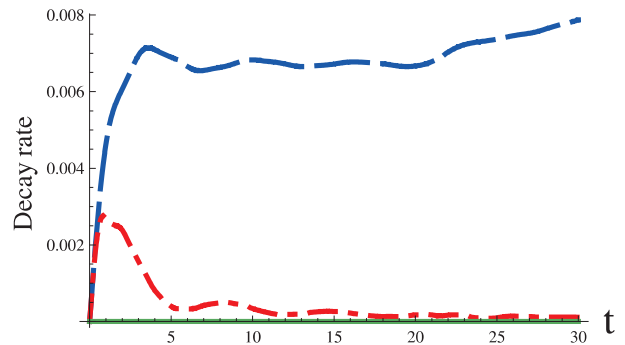}
\caption{(Color online) The decay rate of the TLS as a function of time $t$
with $t$ in units of $\protect\xi ^{-1}$. The coupling strength $g=0.25$ and
the number of cavities $N=41$, $\Delta =1,A/\protect\nu =1$ for blue dashed
line, $\Delta =3,A/\protect\nu =1$ for red dot-dashed line, $\Delta =3,A/%
\protect\nu =2.4$ for green solid line. All parameters are in units of $%
\protect\xi $.}
\label{fig4:1}
\end{figure}

We now consider the dynamics of the periodically driven TLS with a finite
time. The investigation starts with Eq.(\ref{ZE-09}). Obviously, the
decoupling between the TLS and the CRW still happens when the ratio of the
driven intensity $A$ to frequency $\nu $ meets a root of the Bessel function
$J_{n}\left( x\right) $. This decoupling preserves the population of the
TLS. Consequently, QZE appears in this system. For $J_{n}\left( A/\nu
\right) \neq 0$, one can still have QZE because the decay rate $R$ grows as $%
t$ increases when equation (\ref{ZE-11}) is satisfied. However, When the
transition energy $\Omega $ is out of resonance with the energy band of the
CRW for any $k$, i.e. equation (\ref{ZE-11}) is not satisfied for any $k$,
the decay rate is roughly a descending function of $t$. Consequently, the
AZE occurs. Hence, for a given driven source, the Zeno-Anti-Zeno crossover
can be adjusted by the detuning between the TLS and the CRW. For a given
detuning $\Delta $, one can switch the occurrence of the QZE and AZE by
controlling the driven intensity and frequency. In Fig.\ref{fig4:1}, we plot
the decay rate as a function of time $t$ with the coupling strength $%
g=0.25\xi $ and the number of the resonators $N=41$. For the green solid
line, the detuning $\Delta =3$, the ratio of the intensity to frequency $%
A/\nu =2.4$. which satisfies the $J_{0}(A/\nu )=0$, therefore the decay rate
is vanished. For the red dot-dashed line, the transition energy of the TLS
is outside the band of the CRW, hence, anti-Zeno effect appears. For the
blue dashed line, transition energy inside the band, consequently, quantum
Zeno effect occurs. Therefore, one can switch quantum Zeno effect to
Anti-Zeno effect by tuning the transition energy of the TLS or the intensity
and frequency of the driven field.

As for the case with infinite cavities and finite time, the important
factors are the the width and center of the spectrums, which determine the
appearance of Zeno and anti-Zeno effect. We denote th width and center of
the modulation spectrum as $\Delta _{f}=t^{-1}$ and $\omega _{f}=\Delta
+n\nu $. The width and center of the reservoir coupling spectrum $%
g_{n}\left( \omega \right) $ read $\Delta _{g}=\sqrt{2}\xi gJ_{n}\left(
A/\nu \right) $ and $\omega _{g}=0$, respectively. When $\Delta _{f}\gg
\Delta _{g},\omega _{f}$, the effective decay rate $R\sim 2\pi f_{n}\left(
\Delta +n\nu \right) $, which grows with $t$, consequently, the quantum Zeno
effect generally occurs. When $\Delta _{f}\ll \left\vert \omega _{f}-\omega
_{g}\right\vert $ and $\omega _{f}$ is significantly detuned from the center
$\omega _{g}$ of $g_{n}\left( \omega \right) $, the effective decay $R\sim
2\pi g_{n}\left( \Delta +n\nu \right) \ll 2\pi g_{n}\left( \omega
_{g}\right) $. As a result, the effective decay $R$ is a increasing function
of the width $\Delta _{f}$, which leads to the acceleration of decay, i.e.
the anti-Zeno effect.

\section{Conclusion}

In summary, we have considered a CRW with a period-driven TLS inside one of
the resonators. The spontaneous emission of the TLS is investigated in the
one-quantum subspace via the Floquet representation of this system. Via
computing the Floquet-Green's function up to the second nonvanishing order
in the coupling strength, the amplitude for the TLS still in its excited
state is obtained. It is found that the band structure of the CRW in the
dispersion relation of the electromagnetic field gives rise to crossover of
the quantum Zeno effect and anti-Zeno effect, which can be controlled by the
driven intensity $A$ and frequency $\nu $, the detuning $\Delta$ between the
cavity and the TLS. One important effect induced by the diagonal couplings
between the driving field and the TLS is the strong dynamical suppression of
decaying at suitable values of applied ac field.

This work is supported by the Program for New Century Excellent Talents in
University (NCET-08-0682), NSFC No.~10775048, No.~10704023, and No.
11074071, NFRPC 2007CB925204, PCSIRT No.~IRT0964, the Project-sponsored by
SRF for ROCS, SEM [2010]609-5, the Key Project of Chinese Ministry of
Education (No.~210150), and Scientific Research Fund of Hunan Provincial
Education Department No.~09B063, and No.~09C638.\vspace*{-0.1in}\vspace*{%
-0.1in}

\appendix

\section{Floquet formulation}

Here, we give a brief outline of this method. According to Floquet's theorem%
\cite{Floquet,Shirley,SIchupr}, the time-dependent Schr\"{o}dinger equation
exists Floquet-state solutions of the form%
\begin{equation}
\left\vert \Psi _{\alpha }\left( t\right) \right\rangle =e^{-i\varepsilon
_{\alpha }t}\left\vert \Phi _{\alpha }\left( t\right) \right\rangle ,
\label{rep-1}
\end{equation}%
where the Floquet state $\left\vert \Phi _{\alpha }\left( t\right)
\right\rangle =\left\vert \Phi _{\alpha }\left( t+T\right) \right\rangle $
is a function with the same period as the driving field, $\varepsilon
_{\alpha }$ is a real-valued energy function and is called the quasi-energy.
The Floquet states obey the eigenvalue equation%
\begin{equation}
\left( H-i\frac{\partial }{\partial t}\right) \left\vert \Phi _{\alpha
}\left( t\right) \right\rangle =\varepsilon _{\alpha }\left\vert \Phi
_{\alpha }\left( t\right) \right\rangle .  \label{rep-2}
\end{equation}%
The operator on the left hand of the above equation gives the Floquet
Hamiltonian
\begin{equation}
H_{F}\equiv H-i\frac{\partial }{\partial t},  \label{rep-3}
\end{equation}%
which is time-independent. Floquet Hamiltonian operates on an extended
Hilbert space $\mathcal{R}=\mathcal{H}\otimes \mathcal{T}$, which made up of
the Hilbert space $\mathcal{H}$ of the system and the temporal space $%
\mathcal{T}$ of time-periodic functions. The temporal part can be spanned by
the orthonormal set of functions $\left\langle t\right. \left\vert
m\right\rangle =\exp \left( im\nu t\right) $, where $m=0,\pm 1,\pm 2,\cdots $%
is the Fourier index, and%
\begin{equation}
\left\langle n\right. \left\vert m\right\rangle =\frac{1}{T}\int_{0}^{T}\exp %
\left[ i\left( m-n\right) \nu t\right] dt=\delta _{mn}.  \label{rep-4}
\end{equation}%
The composite Hilbert space $\mathcal{R}$ is called Floquet or Sambe space%
\cite{Shirley}. In Floquet space, the Floquet Hamiltonian is linear and
Hermitian, and the Floquet states provide a complete basis with the scalar
product defined as%
\begin{equation}
\langle \langle \Phi _{\alpha }\left( t\right) \mid \Phi _{\beta }\left(
t\right) \rangle \rangle =\frac{1}{T}\int_{0}^{T}\left\langle \Phi _{\alpha
}\left( t\right) \right. \left\vert \Phi _{\beta }\left( t\right)
\right\rangle dt.  \label{rep-5}
\end{equation}%
The matrix elements of the time-evolution operator $U_{\alpha \beta }\left(
t,t_{0}\right) $ propagates the state $\left\vert \alpha \right\rangle $ at
time $t_{0}$ to the state $\left\vert \beta \right\rangle $ at time $t>t_{0}$
according to the time-dependent Hamiltonian $H$. In\ the Floquet
representation, $U_{\alpha \beta }\left( t,t_{0}\right) $ is related to the
Floquet Hamiltonian%
\begin{equation}
U_{\alpha \beta }\left( t,t_{0}\right) =\sum_{n}\left\langle \beta
n\right\vert \exp \left[ -iH_{F}\left( t-t_{0}\right) \right] \left\vert
\alpha 0\right\rangle e^{in\omega t}  \label{rep-6}
\end{equation}%
and is interpreted in Ref.\cite{Shirley} as "the amplitude that a system
initially in the Floquet state $\left\vert \alpha 0\right\rangle $ at time $%
t_{0}$ evolves to the Floquet state $\left\vert \beta n\right\rangle $ at
time $t$ according to the time-independent Floquet Hamiltonian $H_{F}$,
summed over $n$ with weighting factors $\exp \left( in\omega t\right) $". In
experiment, the probability to go from the initial state $\left\vert \alpha
\right\rangle $ to the final state $\left\vert \beta \right\rangle $ is the
time-averaged transition probability between $\left\vert \alpha
\right\rangle $ and $\left\vert \beta \right\rangle $%
\begin{equation}
P_{\alpha \beta }=\sum_{n}\left\vert \left\langle \beta n\right\vert \exp
\left[ -iH_{F}\left( t-t_{0}\right) \right] \left\vert \alpha 0\right\rangle
\right\vert ^{2}\text{.}  \label{rep-7}
\end{equation}%
The matrix elements $U_{\alpha \beta }\left( t,t_{0}\right) $ is related to
the Floquet-Green's functions via the Cauchy integral formula $U_{\alpha
\beta }\left( t,t_{0}\right) =\oint e^{-iEt}G_{\alpha \beta }dE$ with the
Floquet-Green's functions\cite{luisPRB}%
\begin{equation}
G_{\alpha \beta }=\sum_{n}e^{in\omega t}G_{\alpha \beta }^{[n]}\text{.}
\label{rep-8}
\end{equation}%
The right hand of equation (\ref{rep-8}) is the Fourier expansion of $%
G_{\alpha \beta }$ with the Fourier coefficients
\begin{equation}
G_{\alpha \beta }^{[n]}=\left\langle \beta n\right\vert \left(
E-H_{F}\right) ^{-1}\left\vert \alpha 0\right\rangle \text{.}  \label{rep-9}
\end{equation}

\end{document}